# Efficient Charge-Spin Conversion and Magnetization Switching though Rashba Effect at Topological Insulator/Ag Interface

Shuyuan Shi[1†], Aizhu Wang[1,2†], Yi Wang[1], Rajagopalan Ramaswamy[1], Lei Shen[3], Jisoo Moon[4], Dapeng Zhu[1], Jiawei Yu[1], Seongshik Oh[4], Yuanping Feng[2], and Hyunsoo Yang[1]*

[1]Department of Electrical and Computer Engineering, National University of Singapore, 117576, Singapore

[2]Department of Physics, National University of Singapore, 117542, Singapore

[3]Department of Mechanical Engineering & Engineering Science Programme, National University of Singapore, Singapore 117575, Singapore.

[4]Department of Physics and Astronomy, Rutgers, The state University of New Jersey, Piscataway, NJ 08854, USA

We report the observation of efficient charge-to-spin conversion in the three-dimensional topological insulator (TI) $Bi_2Se_3$ and Ag bilayer by the spin-torque ferromagnetic resonance technique. The spin orbit torque ratio in the $Bi_2Se_3$/Ag/CoFeB heterostructure shows a significant enhancement as the Ag thickness increases to ~2 nm and reaches a value of 0.5 for 5 nm Ag, which is ~3 times higher than that of $Bi_2Se_3$/CoFeB at room temperature. The observation reveals the interfacial effect of $Bi_2Se_3$/Ag exceeds that of the topological surface states (TSS) in the $Bi_2Se_3$ layer and plays a dominant role in the charge-to-spin conversion in the $Bi_2Se_3$/Ag/CoFeB system. Based on the first-principles calculations, we attribute our observation to the large Rashba-splitting bands which wrap the TSS band and has the same net spin polarization direction as TSS of $Bi_2Se_3$. Subsequently, we demonstrate for the first time the Rashba induced magnetization switching in $Bi_2Se_3$/Ag/Py with a low current density of $5.8 \times 10^5$ $A/cm^2$.

Manipulating the magnetization through spin orbit torques (SOTs) is one of the primary research fields in spintronics [1-4]. A key objective is to efficiently convert charge to spin currents. The spin Hall effect (SHE), as a mechanism to realize charge-to-spin conversion, has been extensively studied in heavy metals [2,5,6]. However, due to the bulk nature of SHE, a certain thickness of heavy metals is required to achieve a sufficient spin current density. Recently, Rashba induced charge-spin interconversions, with a higher efficiency than that from SHE, have been detected in Bi/Ag, $\alpha$-Sn/Ag [7-10], $LaAlO_3/SrTiO_3$ [11]. On the other hand, the topological surface states (TSS) of topological insulators (TIs) are another interfacial mechanism for achieving efficient charge-to-spin conversion due to the strong spin orbit coupling (SOC) and inherent spin-momentum locking, despite of the inevitable bulk transport [3,4,12-17]. Therefore, understanding and optimization of the interfacial effects is of significance and great interest for efficient spin-current generation in SOT based spintronics devices.

Angle-resolved photoemission spectroscopy (ARPES) studies and theoretical analyses have confirmed the coexistence of TSS and Rashba splitting states (RSS) on TI surfaces [18-23]. Since the two branches of Rashba splitting bands are wrapped by TSS linear bands, it is generally assumed that the Rashba bands play a less role in spin-current generation than TSS [24]. In addition, the effective spin polarization of the Rashba splitting bands was found to be opposite to that of TSS, which may cause a partial cancellation of the spin polarizations from TSS and thus a reduction of SOT from TI surfaces [25,26]. However, recent ARPES measurements reveal that the strength of Rashba splitting on the three-dimensional (3D) TI, $Bi_2Se_3$, surface can be tuned by involving adsorbates, such as Pd and Cs [19,21]. Moreover, a recent theoretical work suggests that a new kind of large Rashba splitting states is expected to appear particularly at the $Bi_2Se_3$/Ag interface induced by the spin chirality of TSSs, which possess stronger spin polarizations than TSS in $Bi_2Se_3$ [27]. Therefore, this emerging RSS are promising to generate spin currents more efficiently. However, there has been no experimental demonstration and evaluation of the charge-to-spin conversion at the $Bi_2Se_3$/Ag interface yet. Moreover, the spin configurations as well as the interactions between TSS and RSS have not been clearly identified in the TI based systems.

In this Letter, we study the spin current generation in $Bi_2Se_3$/Ag/CoFeB with different Ag thicknesses by the spin-torque ferromagnetic resonance (ST-FMR) technique [4,6,28] at room temperature and observe an enhancement of charge-to-spin conversion efficiency as the thickness of Ag increases. Further analyses indicate that the efficient charge-to-spin conversion is primarily a result of the Rashba effect at the $Bi_2Se_3$/Ag interface. The first-principles calculations reveal the Rashba band structures of the $Bi_2Se_3$/Ag interface are located outside the TSS band and the sign of the spin polarization in the Rashba bands is the same as TSS. Thus, our findings reveal novel physical phenomena on the TI surface and demonstrate where the RSS can contribute to the charge-to-spin conversion together with TSS in TI based spintronics devices.

Film structures of $Al_2O_3$ substrate/$Bi_2Se_3$ (10 QL)/Ag ($t_{Ag}$)/$Co_{40}Fe_{40}B_{20}$ (CoFeB, 7 nm)/MgO (2 nm)/$SiO_2$ (4 nm) (QL: quintuple layer) and control samples of $Al_2O_3$ substrate/Ag ($t_{Ag}$)/CoFeB (7 nm)/MgO (2 nm)/$SiO_2$ (4 nm) were prepared, where $t_{Ag}$ = 0, 1, 2, 3, and 5 nm. The $Bi_2Se_3$ films were grown by molecular beam epitaxy [29,30]. The atomic force microscopy (AFM) image in Fig. 1(a) shows a $Bi_2Se_3$ surface with a roughness of ~0.5 nm with uniformly distributed triangular terraces with a step height of ~1.0 nm, indicating the three-fold symmetry of the $Bi_2Se_3$ (111) plane [29]. The resistivity as a function of temperature in a $Bi_2Se_3$ thin film (10 QL) capped with MgO (2 nm)/$SiO_2$ (4 nm) is shown in Fig. 1(b). The $Bi_2Se_3$ demonstrates a metallic characteristic similar to previous reports [31,32].

To carry out ST-FMR measurements, the Ag and CoFeB layers were deposited on $Bi_2Se_3$ using magnetron sputtering with a base pressure $< 3 \times 10^{-9}$ Torr. The films were patterned into microstrips with dimensions of 15–22.5 µm in width and 130 µm in length by photolithography and ion milling. Coplanar waveguides with Ta (2 nm)/Cu (150 nm) were deposited subsequently. The gap between the ground and signal electrodes was varied to tune the device impedance to be ~50 Ω. As indicated in Fig. 1(c), during ST-FMR measurements, a radio frequency (rf) current ($I_{rf}$) with frequencies from 6 to 9 GHz and a power of 15 dBm was applied along the *x* axis using a signal generator. An external magnetic field ($H_{ext}$) was swept in the *x-y* plane with an angle ($\theta_H$) of 38° with respect to the *x* axis. The CoFeB magnetization precesses due to the rf current-induced torques, including the in-plane torque ($\tau_\parallel$, also called damping-like torque) from the oscillating spin current and the out-of-plane torque ($\tau_\perp$) mainly from the oscillating rf current induced Oersted field. The magnetization precession yields the oscillation of the device resistance due to the anisotropic magnetoresistance (AMR) of CoFeB, which is measured as a ST-FMR voltage ($V_{mix}$) by a lock-in amplifier.

A representative ST-FMR signal for a $Bi_2Se_3$ (10 QL)/Ag (2 nm)/CoFeB (7 nm) sample measured at 6 GHz is shown in Fig. 1(d). The ST-FMR signal is the combination of the symmetric ($V_{sym}$) and anti-symmetric ($V_{asym}$) Lorentzian components, which can be separated by fitting the mixed signal with $V_{mix} = V_{sym} F_s(H_{ext}) + V_{asym} F_a(H_{ext})$ [3,4,6,28]. Consequently, $V_{sym}$ (proportional to $\tau_\parallel$) and $V_{asym}$ (proportional to $\tau_\perp$) can be extracted. Figure 1(e) shows the obtained $V_{mix}$ versus $H_{ext}$ spectra from the same sample at different rf frequencies ($f$ = 6, 7, 8, and 9 GHz). The inset of Fig. 1(e) shows *f* as a function of the resonant field $H_0$, which is fitted by the Kittel formula $f = \gamma/2\pi \left[ H_0 \left( H_0 + 4\pi M_{eff} \right) \right]^{1/2}$, confirming the excitation of the ferromagnetic resonance (FMR) in the CoFeB layer [33]. The effective magnetization $M_{eff}$ of CoFeB is determined to be ~1.15 T, which is comparable to the reported value [4].

Figure 2(a) shows the ST-FMR spectra from $Bi_2Se_3$ (10 QL)/Ag (0, 1, 2, 3, and 5 nm)/CoFeB (7 nm). The signal amplitude with 5 nm Ag is small due to a small AMR. As can be seen, the line shape is more symmetric-like with thicker Ag insertion layers, indicating a larger in-plane torque $\tau_\parallel$. Figure 2(b) shows the $\tau_\parallel$ normalized to the current density of $10^8$ A/cm$^2$ for the $Bi_2Se_3$ (10 QL)/Ag ($t_{Ag}$)/CoFeB (7 nm) samples and Ag ($t_{Ag}$)/CoFeB (7 nm) control samples, where $\tau_\parallel$ is the averaged value with $f$ = 6, 7, and 8 GHz. We find that $\tau_\parallel$ for $Bi_2Se_3$/Ag/CoFeB shows a significant enhancement as the Ag thickness increases. We can rule out the influence of the SHE of Ag layer to the increase of $\tau_\parallel$ in the $Bi_2Se_3$/Ag/CoFeB system, as $\tau_\parallel$ in Ag/CoFeB control samples is negligibly small [7,34].

To further explore the role of the $Bi_2Se_3$/Ag interface in the charge-to-spin conversion, we analyze the ratio of in-plane and out-of-plane torque ($\tau_\parallel / \tau_\perp$) [35] for $Bi_2Se_3$/Ag/CoFeB samples. As shown in Fig. 2(c), as we increase $t_{Ag}$, the $\tau_\parallel / \tau_\perp$ increases from 0.7 ($t_{Ag}$ = 0 nm) to 1.7 ($t_{Ag}$ = 2 nm), indicating that the $Bi_2Se_3$/Ag ($t_{Ag}$)/ CoFeB system changes from a $\tau_\perp$ dominating to $\tau_\parallel$ dominating region with increasing $t_{Ag}$. Particularly, an abrupt enhancement appears at $t_{Ag}$ = 2 nm, followed by a saturation behavior, suggesting that the $Bi_2Se_3$/Ag interface effect saturates above 2 nm Ag. This $\tau_\parallel / \tau_\perp$ evolution feature is consistent with the $\tau_\parallel / \tau_\perp$ measured in the Bi/Ag ($t_{Ag}$) interface Rashba system reported recently [35], suggesting that the $Bi_2Se_3$/Ag interface effect is most likely to be the interfacial Rashba effect. We also extracted $\tau_\parallel / \tau_\perp$ ratio in the Ag/CoFeB control samples without the $Bi_2Se_3$/Ag interface in Fig. 2(c). The $\tau_\parallel / \tau_\perp$ ratio in the Ag/CoFeB samples shows a mild increment as $t_{Ag}$ becomes thicker without a saturation behavior, which is in line with the small SHE and long spin diffusion length (~300 nm at room temperature) in Ag [34].

The figure of merit for the charge-to-spin conversion in $Bi_2Se_3$/Ag/CoFeB samples can be characterized as the spin orbit torque ratio ($\theta_\parallel$) defined by the spin Hall conductivity ($\sigma_s$) over charge conductivity ($\sigma$) in a spin current source: $\theta_\parallel = \sigma_s / \sigma$, $\sigma_s = \tau_\parallel M_s t / E$ [3,4,6], where $E$ is the microwave field across the device. $M_s$ and $t$ represent the saturation magnetization and the thickness of the CoFeB layer, respectively. By adoping the established analysis method [3,4], we obtain the $\theta_\parallel$ in $Bi_2Se_3$/Ag/CoFeB samples as shown in Fig. 2(d). The $\theta_\parallel$ is significantly improved with inserting the Ag layer between $Bi_2Se_3$ and CoFeB. Notably, for the $Bi_2Se_3$/Ag (5 nm)/CoFeB device, the $\theta_\parallel$ is ~0.5 while the $\theta_\parallel$ for $Bi_2Se_3$/CoFeB is less than 0.15. Moreover, we find the $\theta_\parallel$ for the Ag/CoFeB control samples is negligibly small compared to the large $\theta_\parallel$ ~ 0.5 obtained in $Bi_2Se_3$/Ag (5 nm)/CoFeB.

We further examine the sign of $\theta_\parallel$, from which we can obtain the spin polarization direction. First, for $Bi_2Se_3$/CoFeB sample ($t_{Ag}$ = 0 nm), the direction of in-plane spin polarization to the electron momentum agrees with spin-momentum locking at TSS [3,4,14,36]. For the samples with $t_{Ag}$ > 0 nm, the interfacial Rashba effect comes into play and, in literature, the spin polarization of the Rashba effect on TI was found to be generally opposite to that of the TSS [25,26]. However, from our measurements, it is clear that accumulated spins due to the $Bi_2Se_3$/Ag interface effect and $Bi_2Se_3$ TSS are in the same direction, as the spin accumulation is greatly enhanced in the $Bi_2Se_3$/Ag/CoFeB samples compared with the $Bi_2Se_3$/CoFeB sample.

In order to understand the underline physics of the efficient charge-spin conversion by inserting the Ag layer between $Bi_2Se_3$ and CoFeB, we study the interfacial Rashba states, topological surface states, and the corresponding spin configurations of $Bi_2Se_3$/Ag ($t_{Ag}$ = 0-3.59 nm)/CoFeB using the first-principles. An energetically favored heterostructure, $Bi_2Se_3$ (1×1)/Ag (1×1), with the selenium termination is adopted [37-39]. The ion-electron, electron-electron and van der Waals (vdW) interactions between Ag and $Bi_2Se_3$ are included in our model implemented in the Vienna *ab* initio simulation package (VASP) [24,40,41]. Figure 3(a) shows the calculated Rashba parameter ($\alpha_R$, the figure of merit of the strength of Rashba splitting) of $Bi_2Se_3$/Ag [39]. As can be seen, $\alpha_R$ increases with $t_{Ag}$ until reaching a saturated value of ~3.07 eVÅ (or $\alpha_R/\hbar$ ~ 4.66 × $10^5$ m/s) at 1.8 nm. The trend is consistent with our experimental results of the $\tau_\parallel/\tau_\perp$ and spin-orbit-torque ratio in Figs. 2(c) and (d). The saturation of $\alpha_R$ at 1.8 nm can be explained by the interfacial electrostatic potential gradient ($\nabla V$, shown in the upper inset of Fig. 3(a)) as the interfacial Rashba splitting is not only related to the atomic SOC of $Bi_2Se_3$, but also the $\nabla V$ induced by the structural inversion asymmetry in the heterostructure [42,43].

As the RSS and the TSS are coexistent in $Bi_2Se_3$/Ag, we also calculate the Fermi velocity ($v_F$, the figure of merit of Dirac bands) of TSS in Fig. 3(a) to check which one governs the efficient charge-spin conversion [39]. As can be seen, $v_F$ is smaller than $\alpha_R/\hbar$ for $t_{Ag}$ > 1.28 nm, indicating the major role of the RSS on the charge-spin conversion over TSS. The minimum $v_F$ at 0.95 nm is due the distortion of the linear TSS by the bulk states when the Dirac cone ($E_D$) is away from the Fermi level ($E_F$) as shown in the lower inset of Fig. 3(a). A similar observation has been reported in $Bi_{1.5}Sb_{0.5}Te_{1.7}Se_{1.3}$/Ag by ARPES measurements [44].

The conventional RSS are known to suppress the TSS because they are located inside TSS and have an opposite spin texture [19,45,46]. In order to understand the enhancement of the charge-spin conversion with the coexistence of the RSS and the TSS in our case, we calculate the spin textures of $Bi_2Se_3$/Ag. Figures 3(b) and 3(c) show the calculated band structures of $Bi_2Se_3$ (6 QL)/Ag (0 nm) and $Bi_2Se_3$ (6 QL)/Ag (0.95 nm), respectively. Without Ag, the surface states of $Bi_2Se_3$ consist of a single Dirac cone and the Fermi level exactly crosses the Dirac cone [47]. By introducing Ag, a new type of Rashba splitting bands appears, in which the spin degenerate point of RSS is located below $E_F$, indicating an *n*-type doping behavior because of the charge transfer from Ag to $Bi_2Se_3$ [48]. The most different feature of this new RSS is that they are located outside of the TSS. The calculated helical spin textures of the heterostructure with $t_{Ag}$ =0 and 0.95 nm at an energy contour 0.05 eV above the Dirac cone are shown in Fig. 3(d) and 3(e), respectively. The outermost concentric Rashba band has the same spin direction with the inside Dirac band, resulting in an additive effect in the spin polarization, thus an enhancement of charge-spin conversion under in-plane charge currents.

The calculation results of $\alpha_R$ is comparable with our measurements. We extract the $\alpha_R$ from our experiments using $\alpha_R = \hbar \lambda_{IEE}/\tau$ [7,11], where $\lambda_{IEE}$ is the inverse Edelstein length, $\tau$ is the spin lifetime. By using the equation $\lambda_{IEE} = \theta_{SHE} l_{sf}$ [10,11], the overall $\lambda_{IEE}$ is obtained to be ~3.1 nm (2.23 nm from RSS, 0.87 nm from TSS) (Supplemental Material [39]), which is higher than 0.1-0.4 nm of pure RSS and 2.1 nm of pure TSS in Bi/Ag [7-9] and *α*-Sn/Ag bilayer [10]. This further evidences that the additive effect of the RSS and TSS can lead to a larger charge-to-spin conversion efficiency. In addition, the $\alpha_R$ at the $Bi_2Se_3$/Ag interface is extracted to be 2.83-3.83 eVÅ from our experiments [39], which is close to the value of 3.07 eVÅ obtained from our first-principles calculations. This value of $\alpha_R$ is similar to that of Bi/Ag [48]. Further, the ratio of TSS and RSS, $v_F/(\alpha_R/\hbar)$ from experiments is ~0.389, which is close to the value of 0.32 obtained from our calculations, indicating the dominant role of RSS over TSS in charge-to-spin conversion.

Therefore, our calculations are in good agreement with our measurements, providing physical insights to the experimental observations.

Subsequently, by utilising the high charge to spin conversion efficiency, we for the first time demonstrate the interfacial Rashba effect driven magnetization switching at room temperature in $Bi_2Se_3$ (10 QL)/Ag (2 nm)/Py (6 nm) by magneto-optic Kerr effect (MOKE) microscopy. Figure 4(a) is a microscopic picture of the device, showing a 15 μm wide $Bi_2Se_3$/Ag/Py channel connected with two electrodes. Figures 4(b) and (c) demonstrate the magnetization of Py is switched from down (light) to up (dark) by applying a current density of $5.8 \times 10^5$ A/cm$^2$. Figures 4(d) and (e) show the magnetization of Py is switched back from up to down by an opposite current pulse. It is observed that the $Bi_2Se_3$/Ag can reduce the switching current density down to $\sim 5.8 \times 10^5$ A/cm$^2$, which is about two orders of magnitude smaller than that in conventional heavy metal/ferromagnet bilayers [1,2,5]. Without the Ag insertion (Supplemental Material [39]), we find the required switching current density in $Bi_2Se_3$/Py is much higher ($\sim 1.32 \times 10^6$ A/cm$^2$), confirming that the $Bi_2Se_3$/Ag bilayer has a higher charge-spin conversion efficiency than single $Bi_2Se_3$ due to the additive effect of the TSS and RSS.

In summary, we have studied the charge-to-spin current conversion at the $Bi_2Se_3$/Ag interface by ST-FMR. A prominent enhancement of $\theta_\parallel$ with increasing $t_{Ag}$ is observed, which reaches a value of ~0.5 for $Bi_2Se_3$ (10 QL)/Ag (5 nm)/CoFeB (7 nm) at room temperature. First-principles calculations reveal that the increment in $\theta_\parallel$ is dominated by the large Rashba bands at the Ag/$Bi_2Se_3$ interface, which are located outside the TSS linear bands with the same effective spin polarization direction as $Bi_2Se_3$ TSS. We also found the additive effect of TSS and RSS can reduce the magnetization switching current density from $1.32 \times 10^6$ to $5.8 \times 10^5$ A/cm$^2$. Our work indicated a new method to design efficient TI based spin orbit torque devices.

This work was supported by the A*STAR's Pharos Programme on Topological Insulators. The first-principles calculations were carried out on the GRC-NUS high-performance computing facilities.

†These authors contributed equally to this work.

*eleyang@nus.edu.sg

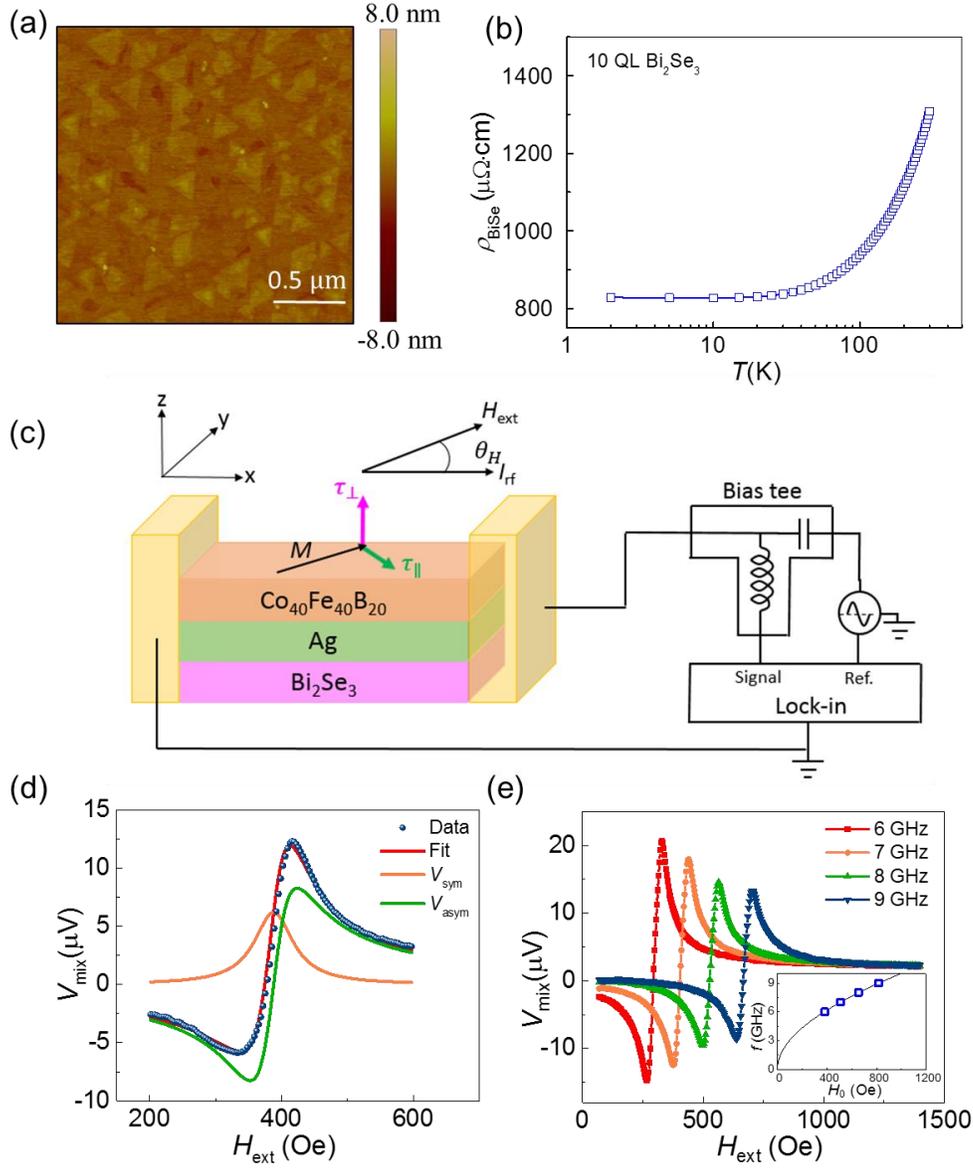

Fig. 1. (a) AFM image of the $Bi_2Se_3$ surface after removing Se cap. (b) Temperature-dependent resistivity of a 10 QL $Bi_2Se_3$ film. (c) Schematic the ST-FMR set-up and the measurement geometry. *M* is the magnetization of the CoFeB layer, which oscillates along an external magnetic field $H_{ext}$ in the *x-y* plane. $\theta_H$ is the angle between $H_{ext}$ and *x*-axis. (d) The ST-FMR signal from a $Bi_2Se_3$ (10 QL)/Ag (2 nm)/CoFeB (7 nm) sample at 6 GHz. The solid lines are fits showing the symmetric ($V_{sym}$) (orange) and anti-symmetric ($V_{asym}$) (blue) Lorentzian contributions. (e) The ST-FMR spectra from the $Bi_2Se_3$/Ag/CoFeB sample at 6–9 GHz. The inset shows the resonance frequency (*f*) as a function of the resonant field ($H_0$) with a line fit to the Kittel formula.

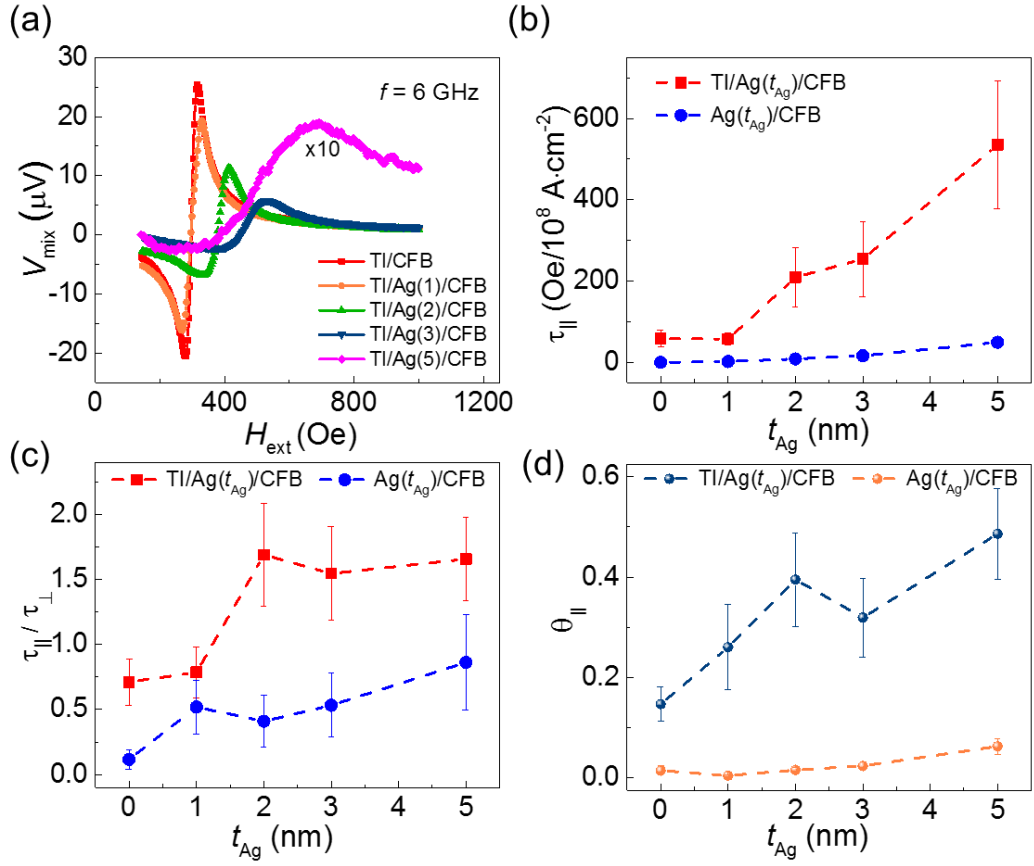

Fig. 2. (a) ST-FMR spectra for TI/Ag ($t_{Ag}$)/CoFeB at 6 GHz. The signal for the Ag (5 nm) sample is scaled by ten times (×10). (b) $\tau_\parallel$ per charge current density without and with the TI layer. (c) $\tau_\parallel/\tau_\perp$ and (d) the spin orbit toque ratio ($\theta_\parallel$) obtained for both TI/Ag ($t_{Ag}$)/CoFeB and Ag ($t_{Ag}$)/CoFeB samples. $Bi_2Se_3$ (10 QL) and CoFeB (7 nm) are denoted as TI and CFB, respectively.

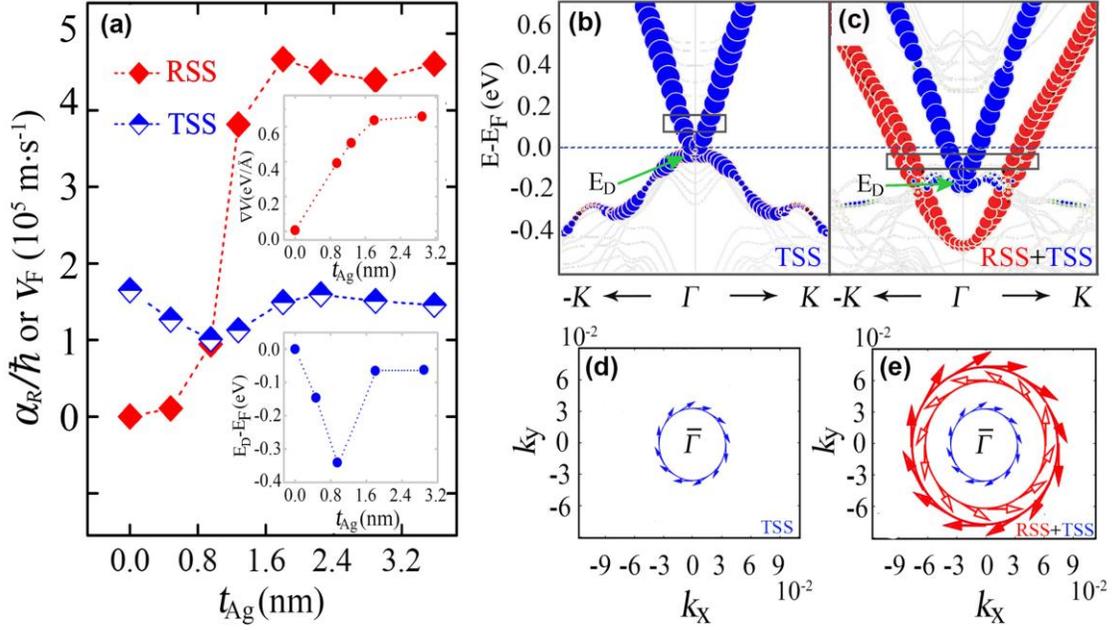

Fig. 3. (a) Calculated evolutions of the strength of RSS ($\alpha_R/\hbar$) and the Fermi velocity of TSS ($v_F$) with respect to $t_{Ag}$. The upper inset is the interfacial electrostatic potential gradient. The below inset shows the energy difference between the Dirac point and the Fermi level. The band structure of (b) $Bi_2Se_3$ (6QL)/Ag (0 nm) and (c) $Bi_2Se_3$ (6 QL)/Ag (0.95 nm). The weight of the first and second quintuple layers is highlighted by the size of circles. The RSS are in red and the TSS are in blue. The helical spin textures in an energy contours of 0.05 eV above the Dirac cone for (d) $Bi_2Se_3$ (6QL)/Ag (0 nm) and (e) $Bi_2Se_3$ (6 QL)/Ag (0.95 nm). The arrows indicate the spin direction within the $k_x$-$k_y$ plane.

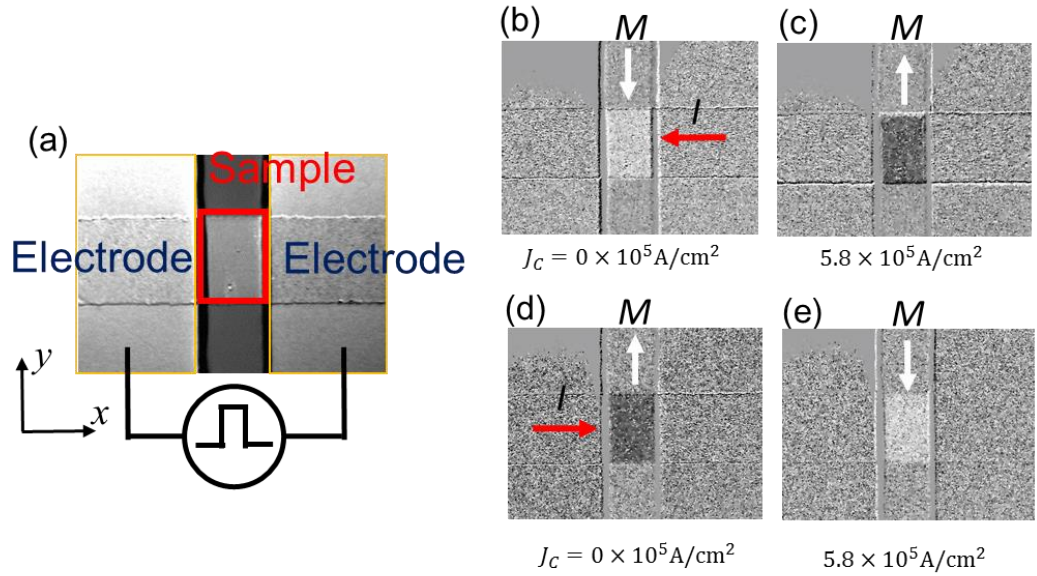

Fig. 4. MOKE images of SOT driven magnetization switching in $Bi_2Se_3$ (10 QL)/Ag (2 nm)/Py (6 nm) at room temperature. (a) The microscopic picture of the device, showing a 15 μm wide $Bi_2Se_3$/Ag/Py channel connected with two electrodes. (b-e) MOKE images for SOT driven magnetization switching by applying a pulsed dc current $I$ (indicated by the red arrow) along the $x$-axis. (b) and (c) show the magnetization is switched from down (light) to up (dark). (d) and (e) show the magnetization is switched back from up (dark) to down (light).